\documentclass[aps,prl,showpacs,twocolumn]{revtex4-1}
\usepackage{float}
\usepackage{amssymb}
\usepackage{amsmath}
\usepackage{graphicx}
\usepackage{stackengine}
\usepackage{esint}
\usepackage{color}
\usepackage{hyperref}

\newcommand{\secl}[1]{\noindent\textit{#1}}

\newcommand{\derx}{\partial_x}

\newcommand{\der}{\partial}

\newcommand{\dd}{\mathrm{d}}

\newcommand{\ii}{\mathrm{i}}

\newcommand{\ie}{\emph{i.e.}\ }

\makeatother

\begin{document}

\title{Full counting statistics and large deviations in thermal 1D Bose gas}

\author{Maksims Arzamasovs}
\affiliation{Department of Applied Physics, School of Science, Xi’an Jiaotong
  University, Xi’an 710049, Shaanxi, China}
\affiliation{Institute of Atomic Physics and Spectroscopy, University of Latvia, Riga, LV-1586, Latvia}
\author{Dimitri M. Gangardt}
\affiliation{School of Physics and Astronomy, University of Birmingham, Edgbaston,
Birmingham, B15 2TT, UK}

\date{\today}

\begin{abstract}
We obtain the distribution of number of atoms in an interval (full
counting statistics) of Lieb-Liniger model of  interacting bosons in one
dimension. Our results are valid in the weakly interacting regime in a
parametrically large window of temperatures and interval lengths. The obtained
distribution deviates strongly from a Gaussian away from the quasi-condensate
regime, and,  for sufficiently short intervals, the probability of large number
fluctuations is strongly enhanced.
\end{abstract}
\maketitle

\secl{Introduction.}
In-situ measurements of particle number fluctuations in a one-dimensional (1D)
ultra cold Bose gas have been recently performed in experiments with ultra
cold $^{87}$Rb atoms on a chip \cite{Esteve2006,ArmijoJacqmin,ArmijoJacqmin2}.
In these experiments absorption images of a 1D gas of interacting bosons are
divided into many intervals of predetermined size $R$ of order of several
microns and the number of atoms in each pixel is inferred based on absorption
intensity. The data accumulated over several repetitions of such imaging was
then used to extract the second \cite{Esteve2006} and third
\cite{ArmijoJacqmin} moments of the obtained particle number
distribution. This distribution, known as full counting statistics (FCS)
contains full information about many-particle correlations. It is also an
object which arises naturally in the experiments
\cite{ArmijoJacqmin,ArmijoJacqmin2}, so it is highly desirable to have
theoretical predictions for FCS.

Despite the fact that one-dimensional bosons with short range interactions are
amenable to description by the exactly solvable Lieb-Liniger model
\cite{LiebLiniger1}, the theoretical treatment of this quantity is a
formidable task \cite{KorepinItsWaldron,Bastianello2018} as it involves
calculation of density correlations between an arbitrary number of different
spatial points \cite{Gaudin,KorepinBogoliubovIzergin}.
It was suggested first in
Ref.~\cite{ArmijoJacqmin} that one can use Yang-Yang thermodynamics of
Lieb-Liniger model \cite{YangYang} if the interval is sufficiently large and
can be viewed as a subsystem in contact with the effective bath characterized
by temperature $T$ and chemical potential $\mu$.  Then the moments of FCS can
be obtained from an appropriate thermodynamic relation involving mean density
of particles $\bar{n}$ as a function of $\mu$ and $T$. This approach was later
extended in Ref.~\cite{Pietraszewicz2017} to calculation of the fourth moment
of FCS.

The results of these studies show that higher moments decay quickly with the
increasing of interval sizes and FCS becomes strongly peaked around mean
number of particles, $\bar{n} R$.  This makes large deviations of particle
number from its mean value extremely improbable. In particular, the emptiness
formation probability, \emph{i.e.} the probability to find a void of size $R$
considered in Ref.~\cite{KorepinItsWaldron,AbanovKorepin,Abanov} is
exponentially small.

The situation is quite the opposite in the limit of microscopic
intervals $\bar{n} R \ll 1$. This limit was recently considered by Bastianello
\emph{et al.}  \cite{Bastianello2018} who obtained FCS using exact analytic
Bethe Ansatz calculations of local multiparticle correlations. An expected
consequence of these studies is that the most probable particle number 
is zero and probability to find  $N$ particles decays as $(\bar{n} R)^N$.

In fact, large deviations of number of particles become appreciable already
for intervals still containing typically a large number of particles,
$\bar{n} R\gg 1$, but shorter than a certain correlation length scale.
For such \emph{mesoscopic} intervals the central limit theorem
does not hold and FCS deviates strongly from the thermodynamic Gaussian
distribution expected for a collection of many  independent  intervals.

\begin{figure}[t]
\includegraphics[width=\columnwidth]{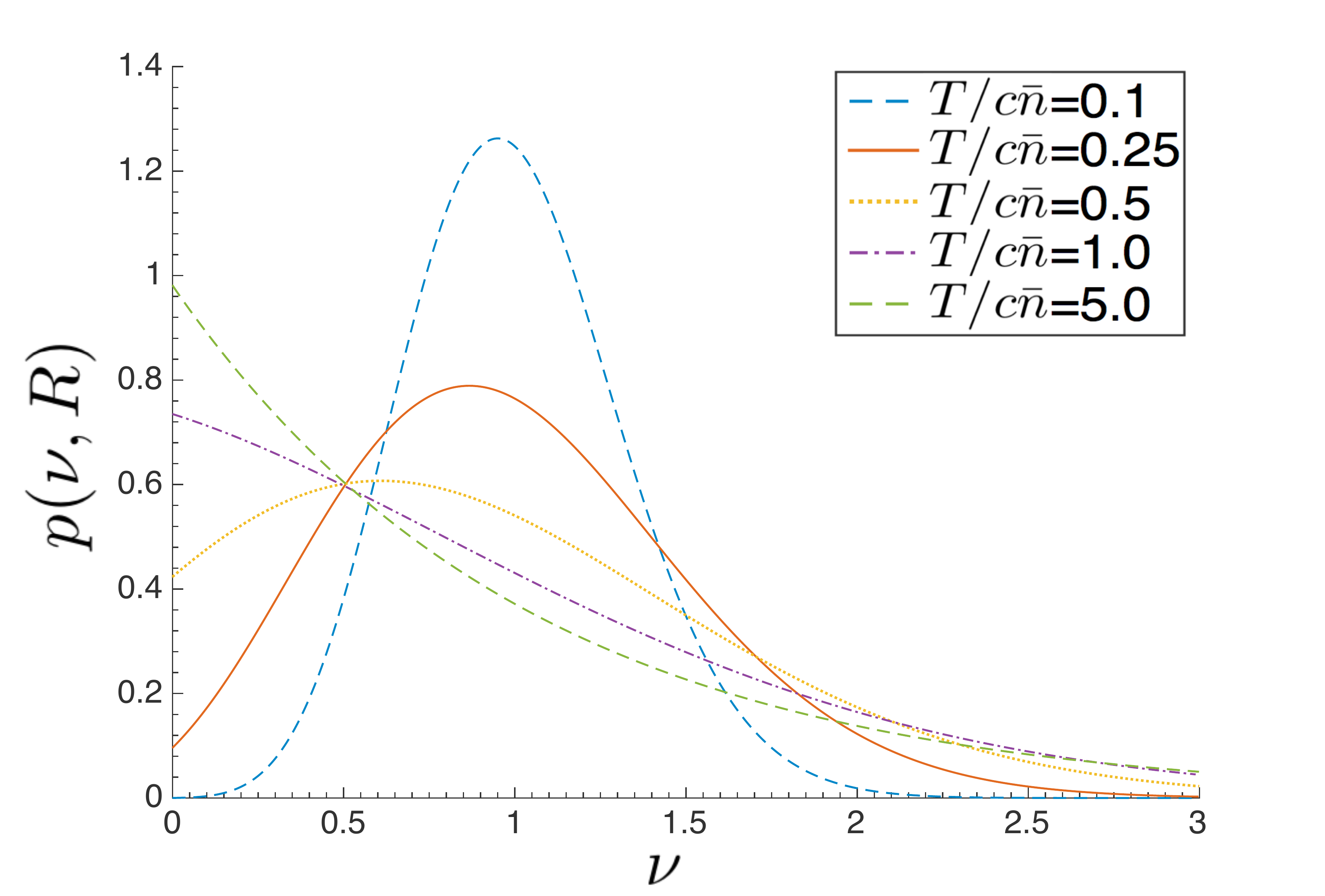}
\caption{Normalized FCS $p(\nu,R)$ defined in Eq.~(\ref{eq:FCS_dens}) 
in the limit of  
short intervals. Dimensionless temperature is 
$T/c\bar{n} = \xi/\ell_\varphi=0.1$, 0.25, 0.5, 1.0, 5.0. 
\label{fig:dist_shortR}}
\end{figure}

\begin{figure}[t]
\includegraphics[width=\columnwidth]{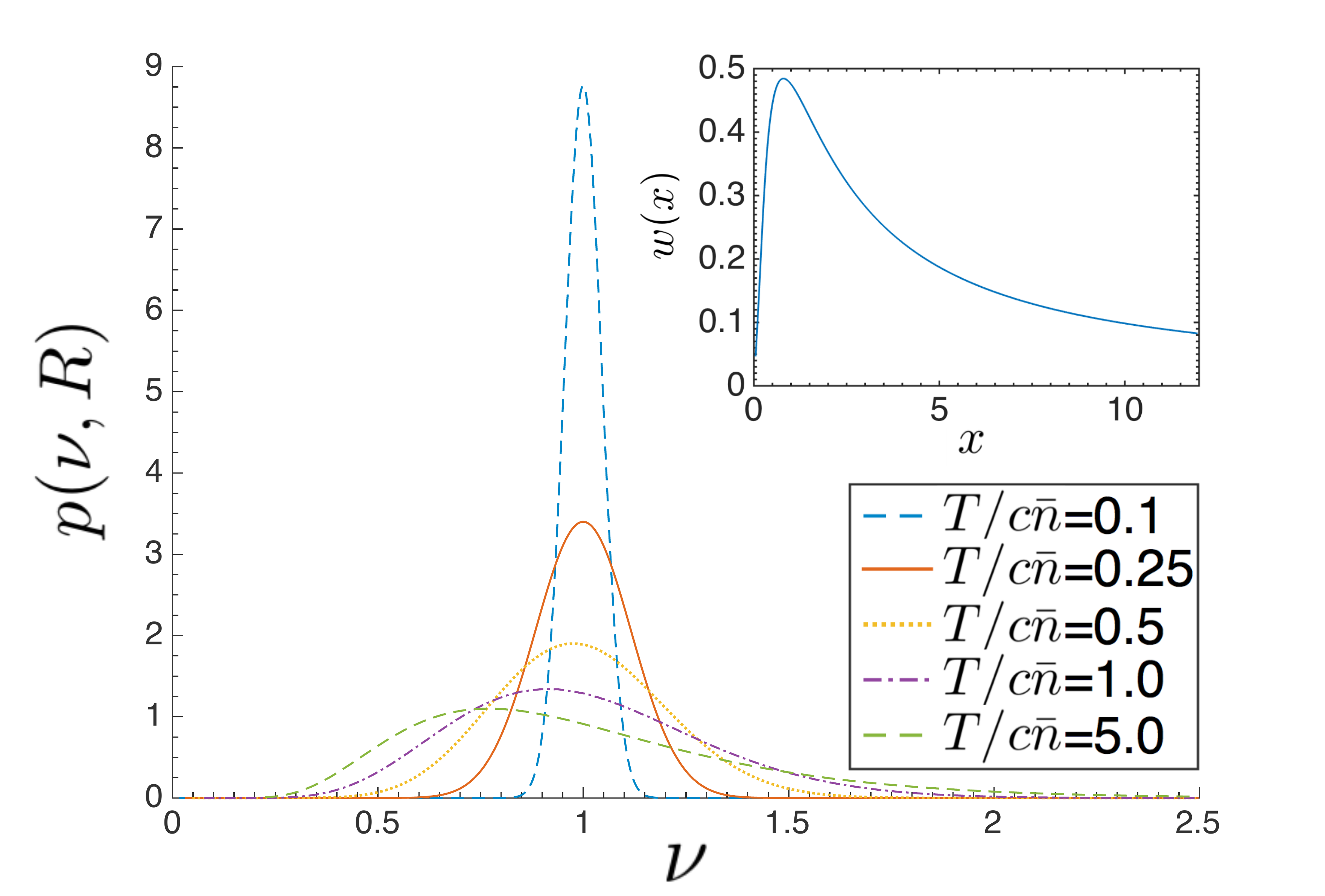}
\caption{Normalized FCS  $p(\nu,R)$ defined in
 Eq.~(\ref{eq:FCS_dens}) 
in the limit of long intervals. Dimensionless temperatures are the same as
in Fig.~\ref{fig:dist_shortR} and $R/\ell_\varphi= 5$. Inset: reduced width 
$w(x)$ defined in Eq.~(\ref{eq:fx}) characterizing fluctuations of number 
of particles. 
\label{fig:dist_longR}}
\end{figure}

\begin{figure}[t]
\includegraphics[width=\columnwidth]{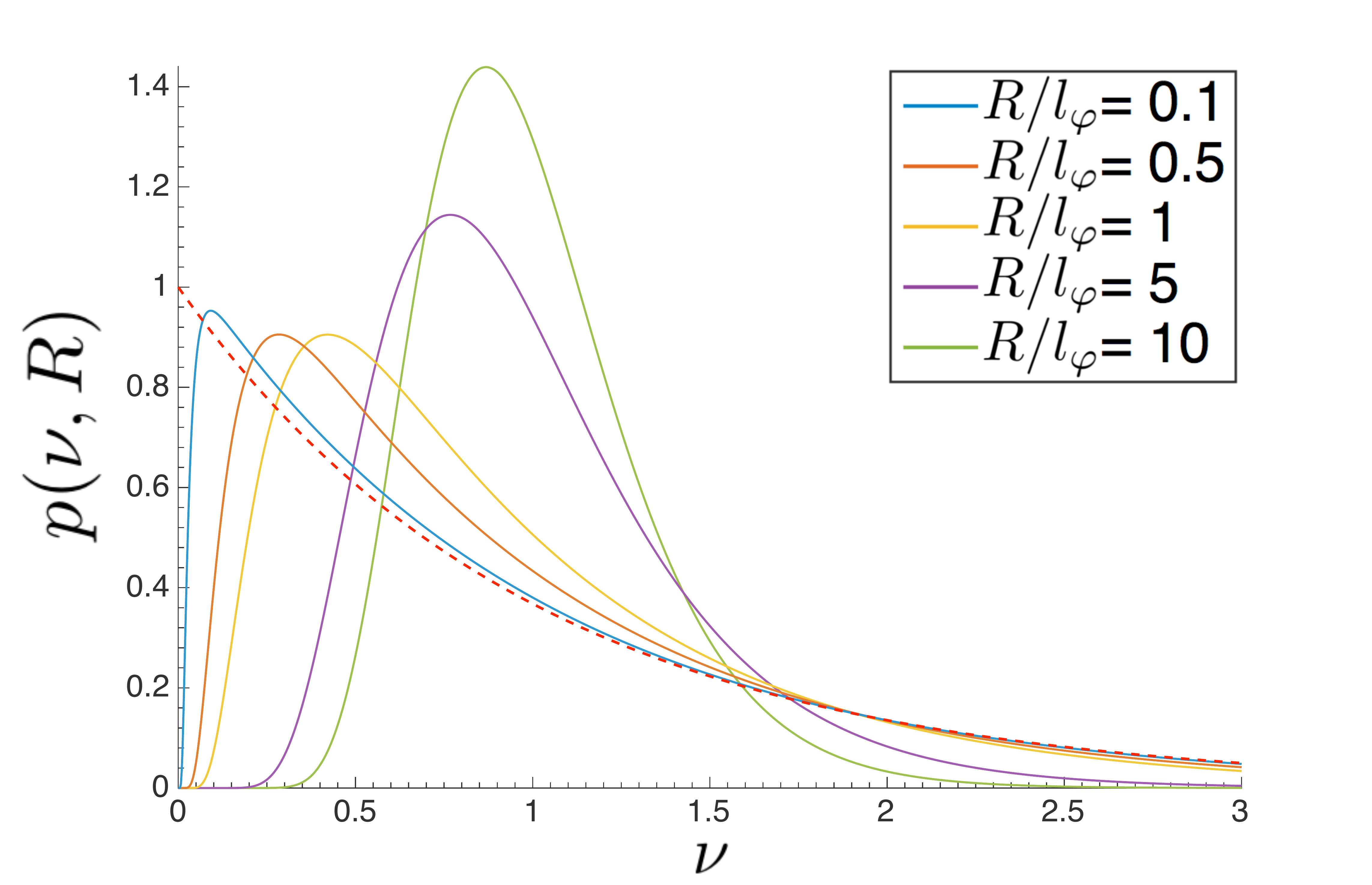}
\caption{Normalized FCS  $p(\nu,R)$ defined in
 Eq.~(\ref{eq:FCS_dens})  
 for intermediate intervals and high temperature regime
 $T/c\bar{n}=\xi/\ell_\varphi  \gg 1$.
 Dashed line represents exponential
distribution, 
Eq.~(\ref{eq:poiss}). 
\label{fig:interR}}
\end{figure}

In this Letter we study FCS on intervals of arbitrary length and provide an
elegant and simple method for its calculation based on the exact mapping of
one-dimensional field theory onto a quantum mechanical problem, introduced in
Ref.~\cite{Castin}. In the limit of short intervals the form of FCS is shown
to change continuously from a Gaussian to an exponential one as temperature is
increased, see Fig.~\ref{fig:dist_shortR}.  The limit of long intervals is
represented in Fig.~\ref{fig:dist_longR} and our FCS agrees with the results
of previous studies.  We also trace FCS as function of the interval length in
Fig.~\ref{fig:interR}.  These results show enhanced large deviations of the
number of particles for mesoscopic intervals where fluctuations play major
role.  

\secl{Full counting statistics.} The main quantity studied in this Letter 
is so-called  full counting statistics
(FCS) defined as  the probability $P_N(R)$ to find exactly $N$ particles in an
interval of length $R$. We define it via the  generating function,
\begin{align}
\label{eq:lt}
  \chi(\lambda,R) &=\sum_{N=0}^\infty e^{-\lambda N} P_N(R) = \left\langle 
e^{-\lambda \hat{N}_R}
\right\rangle\, .
\end{align}
Here $\hat{N}_R =\int_0^R \hat\psi^\dagger\hat\psi\, \dd x$ is the operator of number
of particles in the interval.  
The statistical average in Eq.~(\ref{eq:lt}) is performed in 
the equilibrium state of  uniform 1D Bose gas
with contact interactions. For  normal-ordered operators 
it is given by the imaginary-time functional 
integral
\begin{align}
\label{eq:average}
  \left\langle:\! F[\hat\psi^\dagger,\hat\psi]\!:\right\rangle = \frac{1}{Z}
  \int \mathcal{D}\bar\psi\psi \,F[\bar\psi,\psi]\,e^{-S[\bar\psi,\psi]}\, ,
\end{align}
where configurations of the complex-valued fields 
$\bar\psi(x,\tau),\psi(x,\tau)$ are weighted by the action 
\begin{align}
\label{eq:actionLL}
  S=\int_0^{1/T}\!\dd\tau\!\int\! \dd x \left(\bar\psi\der_\tau\psi+
  \frac{1}{2m}|\der_x\psi|^2-\mu|\psi|^2 
  +\frac{g}{2}|\psi|^4\right)\, , 
\end{align}
with $m$ being the atomic mass and $g$ the
strength of 1D contact interaction. The inverse thermodynamic partition
function $Z$ ensures normalization and the units are chosen such
that $\hbar=1$, $k_B=1$. We are using grand canonical formalism, but    
use  the average density
$\bar{n} = \left\langle \bar\psi(x,\tau)\psi(x,\tau)\right\rangle $
as a control parameter and adjust  chemical potential $\mu$  accordingly.

For intervals containing a large number of particles,  $\bar{n} R \gg 1$, 
it is convenient  to define the distribution 
\begin{align}
\label{eq:FCS_dens}
  p(\nu,R) =\int_{-\ii \infty}^{\ii \infty}\frac{\dd k}{2\pi \ii} e^{k\nu}
  \chi(k/\bar{n} R, R) =
  \bar{n} R P_{\nu \bar{n}R}(R)  
\end{align} 
of the fraction of particles, $\nu = N/\bar{n}R$ treating it as a continuous
variable. We represent the generating function, Eq.~(\ref{eq:lt}), by a
functional integral as it were normal-ordered.  It follows from the relation
$\langle e^{-\lambda \hat N_R} \rangle = \langle :\!e^{(e^{-\lambda} - 1) \hat
  N_R}\! : \rangle$
that the errors introduced by this procedure are of order $1/\bar{n} R$ and we
neglect them.

Here we consider regime where the mean inter-particle separation $1/\bar{n}$ is
the smallest of the characteristic length scales. The other two length
scales, in addition to the interval length, $R$, are  
the healing length $\xi=1/\sqrt{mg\bar{n}}$ and the phase coherence
length $\ell_\varphi = \bar{n}/mT$. Both  are
much longer than $1/\bar{n}$ in the regime of weak interactions $mg/\bar{n}\ll
1$ and degenerate bosons $T<\bar{n}^2/m$.
This leaves only two independent dimensionless
parameters which can be chosen to be $R/\ell_\varphi$ and $\xi/\ell_\varphi$.
The latter equals $T/c\bar{n}$, where $c = \sqrt{g\bar{n}/m}$ is the sound
velocity at zero temperature.  We study FCS as function of these two
parameters.

\secl{Classical field theory and effective quantum mechanics.}  The main
obstacle in calculating FCS is the non linearity of the action
(\ref{eq:actionLL}) which is responsible for correlations between the
particles and which makes the exact calculation of FCS extremely difficult if
at all possible. In the hydrodynamic approach of
Ref.~\cite{AbanovKorepin,Abanov} this difficulty was overcome by expanding
the action (\ref{eq:actionLL}) near configurations of the fields contributing
the most to FCS.  This method is limited to
sufficiently low temperatures, $\xi/\ell_\varphi \ll 1$, and sufficiently
large intervals $R/\xi\gg 1$ where the contribution of quantum and thermal
fluctuations are small.  Here we use an alternative classical field method of
Ref.~\cite{Castin} which accounts properly for thermal fluctuations of
arbitrary magnitude, but not the quantum ones.  The latter can be safely
neglected under condition of sufficiently high temperature, $T\gg g\bar{n}$,
equivalent to $\xi/\ell_\varphi \gg 1/\bar{n}\xi$.  This condition and the
condition of quantum degeneracy $\xi/\ell_\varphi\ll \bar{n}\xi $ define a
parametrically wide range of temperatures where classical field method
provides reliable results for macroscopic intervals of \emph{any length} thus
extending the validity domain of the hydrodynamic approach of
Refs.~\cite{AbanovKorepin,Abanov}.

Neglecting quantum fluctuations amounts to retaining only $\tau$-independent
configurations of fields in Eq.~(\ref{eq:actionLL}), leading to a $1+0$
dimensional field theory described by the action
\begin{align}
\label{eq:seff1}
S\simeq   S_\mathrm{cl} = \frac{1}{T} \int\dd x
  \left(
  \frac{1}{2m}|\derx\psi|^2-\mu|\psi|^2
  +\frac{g}{2}|\psi|^4\right)\, .  
\end{align}
This action can be reformulated as an effective quantum mechanical problem if
we treat the rescaled spatial coordinate $\bar{n}x$ as an effective imaginary
time.  The components of the complex field
$\psi = \sqrt{\bar{n}} re^{\ii \theta} $  are parametrized by  dimensionless
polar coordinates $(r,\theta)$ of a fictitious quantum particle moving in a
plane with the rotationally symmetric Hamiltonian,
\begin{align}
\label{eq:effpot}
 H_0=-\frac{1}{2M} \nabla^2 -\frac{\mu}{T}r^2+\frac{g\bar{n}}{2T}r^4 
\end{align}
with effective mass $M=\bar{n}\ell_\varphi$.  Due to the infinite extension of
the integration in Eq.~(\ref{eq:seff1}) the effective particle is in the
ground state $|0\rangle$ of the Hamiltonian (\ref{eq:effpot}) for
$\bar{n} x =\pm\infty$.  

The shape of the potential in Eq.~(\ref{eq:effpot}) experienced by the
effective quantum particle is controlled by the value of $\mu/T$ obtained from
the condition $\langle 0 | r^2|0\rangle =1$.
It was shown in Ref.~\cite{Castin} that for low temperatures where
$\xi /\ell_\varphi \ll 1$, the chemical potential is positive, $\mu/T>0$, and the
potential experienced by the effective particle has a characteristic ``Mexican
hat'' shape, with the effective particle localized near the valley
$r \simeq 1$. This temperature range corresponds to the quasi-condensate regime
\cite{GangardtShlyap1,GangardtKheruntsyanRaizen}. 
For high temperatures, $\xi/\ell_\varphi \gg 1$, corresponding to quantum
degenerate regime of Refs.\cite{GangardtShlyap1,GangardtKheruntsyanRaizen}, 
$\mu/T<0$  and the effective particle explores  vicinity 
of the minimum at $r=0$, where the potential
is almost harmonic. 

In the language of effective quantum mechanics 
the generating function (\ref{eq:lt})  has the following meaning. 
The ground state $|0\rangle $ is evolved for imaginary time $\bar{n} R $ by
the modified Hamiltonian $ H_\lambda = H_0 +\lambda r^2 $ resulting in the
modified state $e^{-\bar{n}R H_\lambda }|0\rangle $.  The generating function
$\chi(\lambda,R)$ is then given by the normalized overlap
\begin{align}
\label{eq:overlap}
  \chi(\lambda,R) &
=   
\langle 0|e^{-\bar{n} R(H_\lambda -E_0)}|0\rangle 
\, ,
\end{align}
where $E_0$ is the ground state energy of $H_0$.

\secl{Short intervals.}  We first consider the case of a short interval $R$.  In
this limit the imaginary time evolution of the ground state in
Eq.~(\ref{eq:overlap}) is obtained by a multiplication of the rotationally
symmetric ground state wave function $\langle r| 0 \rangle = \Phi_0 (r)$ by an
exponential factor $ e^{-\bar{n} R \lambda r^2}$ so that
\begin{align}
  \chi(\lambda,R) = 2\pi \int r\dd r\,  e^{-\bar{n} R
  \lambda r^2 } |\Phi_0 (r)|^2\, .
\end{align}
The corresponding probability distribution
\begin{align}
  \label{eq:pnshort}
  p(\nu,R) &=  \int \dd k  \int r\dd r\,  e^{\ii k
  (\nu-r^2) }  |\Phi_0 (r)|^2 \nonumber \\
  &=2\pi\int r\dd r\, \delta\left(\nu-r^2\right)  |\Phi_0 (r)|^2 = \pi
    |\Phi_0(\sqrt{\nu})|^2 \, 
\end{align}
is independent of the interval length $R$
and is proportional to the ground state probability density  of the effective 2D
quantum mechanical problem.

For high temperature, $\xi/\ell_\varphi \gg 1$, the ground state $\Phi_0 (r)$
is that of a two-dimensional harmonic oscillator, which is simply  
$\Phi_0 (r) =e^{-r^2/2}/\sqrt{\pi}$, as $1/M\omega_0 =1$.  Using
Eq.~(\ref{eq:pnshort}) we see immediately that  FCS is exponential,
\begin{align}
  \label{eq:poiss}
  p(\nu,R) = e^{-\nu }\, .
\end{align}

The low temperature limit, $\xi/\ell_\varphi \ll 1$, corresponds to the
quasi-condensate regime. Expanding the Mexican hat shaped potential near the
minimum at $r =  1$ we get an effective one-dimensional harmonic
oscillator,
\begin{align}
  \label{eq:vquadr}
  MV(1+\delta r)\simeq 
  \frac{1}{2} \left(\frac{\ell_\varphi}{\xi}\right)^2\left(-1 +4\delta r^2\right),
\end{align}
with the temperature independent frequency  $\omega = 2/\bar{n}\xi $.
The corresponding ground state wave-function 
\begin{align}
  \label{eq:gsho}
\Phi_0 (\delta r)  = \left(\frac{M\omega}{4\pi^3}\right)^\frac{1}{4}
  e^{-\frac{M\omega}{2}\delta r^2}  
\end{align}
yields the approximate Gaussian distribution
\begin{align}
  \label{eq:gsquasic}
  p(\nu,R) = \sqrt{\frac{\ell_\varphi}{2\pi \xi}}
  e^{-\frac{1}{2}\frac{\ell_\varphi}{\xi}
  \left(\nu-1\right)^2}\, . 
\end{align}
The quadratic approximation (\ref{eq:vquadr}) fails for large deviations
$\nu - 1\sim 1 $ and the corresponding quantum mechanical problem has to be
solved numerically. We find numerically  the ground state of the
Hamiltonian~(\ref{eq:effpot}) and plot the corresponding distributions 
in Fig. \ref{fig:dist_shortR} for several values of
$\xi/\ell_\varphi=T/c\bar{n}$.  The plots show how the exponential
distribution in Eq.~(\ref{eq:poiss}) transforms into the Gaussian distribution
of Eq.~(\ref{eq:gsquasic}) with decreasing temperature.

\secl{Long intervals.}  For sufficiently long  intervals $R$ the
evolution operator in Eq. (\ref{eq:overlap}) becomes a projector 
\begin{align}
  e^{-\bar{n} R (H_\lambda - E_0)} \simeq |\lambda\rangle e^{-\bar{n} R
  \delta E_\lambda}\langle \lambda|\, ,
\end{align}
onto the ground state $|\lambda\rangle$ of the modified Hamiltonian,
$ H_\lambda |\lambda\rangle = (E_0+\delta E_\lambda) |\lambda\rangle$.  
The precise criterion separating long intervals from short ones is thus
$\bar{n} R \, \delta E_\lambda\gg 1$. We rewrite
this condition by extracting the kinetic energy scale and defining
$\Delta (s, \xi/\ell_\varphi)=M\delta E_\lambda $, where
$s= \lambda M$. Using the fact that $M = \bar{n} \ell_\varphi$ we see that the
long interval condition becomes 
\begin{align}
  \label{eq:longint}
\frac{R}{\ell_\varphi} \Delta (s,\xi/\ell_\varphi) \gg 1\, .
\end{align}
Provided this condition is satisfied,  the generating function
(\ref{eq:overlap}) has the following form
\begin{align}
  \chi(\lambda,R) = A(\lambda M,\xi/\ell_\varphi)  
  e^{-\frac{R}{\ell_\varphi} \Delta (\lambda
  M,\xi/\ell_\varphi) }\, , 
\end{align}
where the amplitude $A = 
|\langle \lambda| 0\rangle|^2$ is independent
of the interval length. The same condition (\ref{eq:longint}) allows to find
the distribution  (\ref{eq:FCS_dens}) by the saddle point method,
\begin{align}
\label{eq:plong}
  p(\nu,R) 
& = 
\frac{R}{\ell_\varphi} \int \frac{\dd s}{2\pi\ii} \,
  A(s,\xi/\ell_\varphi) 
e^{ \frac{R}{\ell_\varphi} \left(s\nu-\Delta \left(s,\xi/\ell_\varphi\right)\right) } 
\nonumber \\
&\simeq D(\nu,\xi/\ell_\varphi)
  e^{\frac{R}{\ell_\varphi} \Gamma(\nu,\xi/\ell_\varphi)}  
\end{align}
where Legendre transform  
$\Gamma(\nu,\xi/\ell_\varphi) = s \nu - \Delta(s,\xi/\ell_\varphi)$ and the
prefactor      
\begin{align}
  D(\nu,\xi/\ell_\varphi) = \sqrt{\frac{R}{2\pi \ell_\varphi}} 
  \frac{A(s,\xi/\ell_\varphi)}{ 
\sqrt{\left|\der^2_s \Delta (s,\xi/\ell_\varphi)\right|}}\, .
\end{align}
are calculated at the saddle point obtained from  the condition 
$ \nu = \der_s \Delta(s,\xi/\ell_\varphi)$. 

In the high-temperature limit, $\xi/\ell_\varphi \gg 1$, the
rescaled ground state energy shift and the overlap 
become independent of temperature   
$\Delta(s,0) = \sqrt{1+2s}-1$, $A(s,0) = 4\sqrt{1+2s}
\left(1+\sqrt{1+2s}\right)^{-2}$ and we obtain 
\begin{align}
\label{eq:highTlong}
  p(\nu,R) = \sqrt{\frac{R}{2\pi \ell_\varphi }}\ 
  \frac{4e^{\frac{R}{\ell_\varphi}
  \left(1-\frac{1}{2}\left(\nu+\frac{1}{\nu}
  \right) \right)} 
  }{\sqrt{\nu}\left(1+\nu\right)^2}\, . 
\end{align}
For small deviations $|\nu-1|\ll 1$ this expression for FCS  becomes
a  Gaussian with variance
$\overline{\delta \nu^2} =  \ell_\varphi/R $.

In the low temperature regime, $\xi/\ell_\varphi\ll 1$, to the lowest order,  
the rescaled ground  state energy shift is a quadratic function 
$\Delta(s,\xi/\ell_\varphi) \simeq s-(\xi/\ell_\varphi)^2
s^2/2$ and $A(s,\xi/\ell_\varphi) \simeq 1$, so by
performing Gaussian integration we get
\begin{align}
  \label{eq:quasilong}
  p(\nu,R)  = \sqrt{\frac{R\ell_\varphi}{2\pi \xi^2 }}
  \exp\left[-\frac{R \ell_\varphi}{2\xi^2}(\nu -1)^2\right]
\end{align} 
in full agreement with the hydrodynamic 
result of  Refs.~\cite{Abanov,AbanovKorepin}. The variance is
$\overline{\delta \nu^2} =  \xi^2/ \ell_\varphi R $. 
In Fig.~\ref{fig:dist_longR} the results for  $p(\nu,R)$ 
based on numerical calculations of Eq.~(\ref{eq:plong}) are shown  for
intermediate values of dimensionless temperature $T/c\bar{n}=\xi/\ell_\varphi$.

\secl{Intermediate intervals.}  
In the limiting cases of high and low temperature the probability
distribution, can be obtained for an interval of arbitrary length $R$. The
method is based on exact  evolution of harmonic oscillator
wave functions under time-dependent variation of frequency and external force
\cite{BazZeldovichPerelomov}
as explained in Supplemental Material \cite{Supp}.
For high temperatures, $\xi/\ell_\varphi\gg 1$ , 
FCS is shown in Fig.~\ref{fig:interR}. It 
interpolates between  Eq.~(\ref{eq:poiss}), and Eq.~(\ref{eq:highTlong}) and
has distinctive non-Gaussian shape.

For low temperatures, $\xi/\ell_\varphi \ll 1$, the distribution remains very
close to a Gaussian with variance depending on the interval length,
\begin{align}
  \label{eq:maksim}
p(\nu,R) = \sqrt{\frac{1}{2\pi C(R/\xi) }\frac{\ell_\varphi}{\xi }}
  e^{-\frac{1}{2 C(R/\xi)} \frac{\ell_\varphi}{\xi} (\nu-1)^2}\, .
\end{align}
The crossover 
function $C(x) = (2x-1+e^{-2x})/2x^2$
behaves as  $C(x)\simeq 1$ for $x\ll 1$ and $C(x) \simeq 1/x$ for
$x\gg 1$ and interpolates between  Eqs.~(\ref{eq:gsquasic}) and
(\ref{eq:quasilong}).  This result  
could have otherwise been obtained using hydrodynamical approach of
Ref.\cite{Abanov} with gradient terms included in the action.

\secl{Variance of the particle number.}
For a macroscopic  interval, Eq.~(\ref{eq:longint}),
the above results suggest the following scaling form for 
the variance of the number of particles, 
\begin{align}
\label{eq:fx}
  \frac{\overline{\delta N^2}}{\bar{n}R}  = \bar{n} R\overline{\delta \nu^2} 
=  \bar{n}\xi\,  w(\xi/\ell_\varphi)\, ,   
\end{align} 
where the universal  function has the limiting behavior $w(x)= x$
for $x\ll 1$ and $w(x)= 1/x$ for $x\gg 1$.  For intermediate values of
$x$ 
the numerical results for $w(x)$ are shown
in inset in Fig.~\ref{fig:dist_longR} and confirm the non-monotonic dependence
of the particle number variance on temperature anticipated from the limiting
behaviors of $w(x)$. The right hand side of Eq.~(\ref{eq:fx}) is greater than
1 in  the whole range of validity of our approach, $1/\bar{n}\xi <
\xi/\ell_\varphi < \bar{n}\xi$, and thus the fluctuations of particle number
are super-Poissonian in agreement with findings of Ref.~\cite{ArmijoJacqmin2}. 

Higher moments of FCS can also be obtained  from the
knowledge of generating function $\chi(\lambda,R)$ and we  
calculate the third and the fourth
moments in Supplemental Material. They  are in full agreement with the
results of previous studies Refs.\cite{ArmijoJacqmin,Pietraszewicz2017}.

\secl{Concluding remarks.}  The departure of FCS from Poisson distribution
expected for a classical ideal gas \cite{Supp}
is a direct consequence of quantum  statistics and is
closely related to bosonic bunching. At high enough temperatures we found
another manifestation of these quantum effects which lead to an enhanced
probability to find large (on the scale of mean inter-particle separation)
regions of depleted number of particles. 
For lower temperatures the inter-particle interactions
tend to suppresses such large  density deviations from its mean value. 
Our findings are relevant for temperatures, interactions and interval
lengths used in current experiments and can provide a novel way to
characterize the temperature and interaction strength due to the strong
dependence of FCS on these parameters.

\secl{Acknowledgments.} MA would like to thank Prof. Bo Liu for encouragement. 
DMG is grateful to IBS PCS, Daejeon, South Korea, 
for hospitality.

\bibliography{FCSBIB}


\setcounter{equation}{0}
\setcounter{figure}{0}
\setcounter{table}{0}
\setcounter{page}{1}
\makeatletter
\renewcommand{\theequation}{S\arabic{equation}}
\renewcommand{\thefigure}{S\arabic{figure}}
\renewcommand{\bibnumfmt}[1]{[S#1]}
\renewcommand{\citenumfont}[1]{S#1}


\begin{widetext}
\section*{Supplementary material for \\
  ``Full counting statistics and large
  deviations in thermal 1D Bose gas''}
\end{widetext}

\section{Full counting statistics of non-degenerate ideal Bose gas\label{sec:Generalities}}

Full counting statistics for classical
(Boltzmann) non-interacting gas (e.g. see the review of stochastic methods by
Chandrasekhar \cite{Chandrasekhar1}) follows from combinatorial arguments and is given by the binomial
distribution
\begin{equation}
P_{N}(R)=\frac{N_\mathrm{tot}!}{N!(N_\mathrm{tot}-N)!}\left(\frac{R}{L}\right)^{N}\left(1-\frac{R}{L}\right)^{N_\mathrm{tot}-N}\label{eq:BinomialDistribution}
\end{equation} where $N_\mathrm{tot}$ is the total number of particles confined to the system of size $L$ and
held at temperature $T$. In thermodynamic limit,
$N_\mathrm{tot}\rightarrow\infty$, $L\rightarrow\infty$,
and $\bar{n}=const$, the distribution Eq. (\ref{eq:BinomialDistribution})
converges to Poisson distribution,
\begin{equation}
P_{N}(R)\rightarrow\frac{\left(\bar{n}R\right)^{N}e^{-\bar{n}R}}{N!}.\label{eq:BinomialTendsPoisson}
\end{equation}
In the grand canonical case the distribution is still given by
Eq.~(\ref{eq:BinomialTendsPoisson}), 
where the equation of state of classical gas has to be specified,
\[
\bar{n}\lambda=e^{\mu/T},
\]
where $\lambda$ is the
thermal (de Broglie) wavelength,
\begin{equation}
\lambda=\sqrt{\frac{2\pi}{mT}}.\label{eq:deBroglieWavelength}
\end{equation}
and $\mu$ is chemical potential.
This results in the average and variance
\begin{equation}
\overline{N}=\overline{\delta N^{2}}=\overline{N^{2}} -{\overline{N}}^{2}= \bar{n}R.\label{eq:PoissonAverageVariance}
\end{equation}

For $\bar{n}R>1$ the distribution Eq.~(\ref{eq:BinomialTendsPoisson}) has a
maximum at non-zero $N$ and the distribution looks almost Gaussian: going from
$N$ to $\nu=N/\bar{n}R$ and treating it as continuous quantity, obtain
\begin{equation}
  p(\nu,R) \sim
  \exp(-\bar{n}R+\nu\bar{n}R-\nu\bar{n}R\ln(\nu)).\label{eq:PoissonDensity}
\end{equation}
The maximum of the above expression occurs when $\nu=1$ and for $\nu\simeq1$
obtain
\begin{equation}
p(\nu,R)=\sqrt{\frac{\bar{n} R}{2\pi}}e^{-\frac{\bar{n}R}{2}(\nu-1)^{2}}.\label{eq:PoissonGaussian}
\end{equation}
In real systems the above results are relevant in the regime $T>T_d$, where
$T_d=\bar{n}^2/m$ is the temperature of quantum degeneracy, regime beyond the
reach of classical field approximation and therefore not considered in the
main text. 

Extending the above analytic result to degenerate non-interacting Bosons is non-trivial because of correlations caused by quantum statistics. This corresponds to the regime $c\bar{n}<T<T_d$. The crossover from degenerate quantum (corresponding to what is called high temperature regime in this Letter) to classical (corresponding to gas of classical particles discussed above) regimes is manifested by considering the variance of particle number (also see Ref.~\cite{GangardtKheruntsyanRaizen1}). The variance is given by
\[
\overline{\delta N^{2}} =\overline{N}+\int_{0}^{R}\int_{0}^{R}g(x-y)g(y-x)\,\dd x\dd y,
\]
where $g(x-y)=\left\langle \psi^{\dagger}\left(x\right)\psi\left(y\right)\right\rangle $
is one-particle correlation function
\[
g(x-y)=\frac{1}{\sqrt{\pi}\lambda}\int_{-\infty}^{\infty}\frac{\exp\left(2\ii\sqrt{\pi}w\left(y-x\right)/\lambda\right)}{z^{-1}e^{w^{2}}-1}\,\dd w
\]
where $z=e^{\mu/T}$ stands for fugacity. For $T>T_{d}$, $z\ll1$ and the Bosonic
occupation numbers can be replaced by Boltzmann factors, $1/\left(z^{-1}e^{w^{2}}-1\right)\approx ze^{-w^{2}}$.
This leads to the following expression for variance,
\[
\overline{N^{2}} \approx\overline{N} +\overline{N}\sqrt{\frac{T_{d}}{T}}\left\{ \frac{e^{-x^{2}}-1+\sqrt{\pi}x\,\mathrm{erf}\left(x\right)}{x}\right\} ,
\]
where $x=\overline{N}\sqrt{T/T_{d}}$ and the expression
in curly brackets takes values between $0$ and $\sqrt{\pi}$.
In other words, for $T>T_{d}$ the variance is essentially Poissonian
for all interval sizes. 

On the other hand, for $T<T_{d}$, $z\approx1$ and the bosonic occupation
number can be approximated by $1/\left(z^{-1}e^{w^{2}}-1\right)\approx1/\left(1-z+w^{2}\right)$
which leads to
\[
\overline{\delta N^{2}} \approx\overline{N} +\overline{N} ^{2}\left\{ \frac{ 2x-1+e^{-2x}}{2x^{2}}\right\} ,
\]
where $x=\overline{N} /\bar{n}l_\varphi = R/\ell_\varphi$
and the expression in curly brackets takes values ranging from $0$
(for $x\gg1$) to $1$ (for $x=0$). Thus in the quantum degenerate regime
and for short intervals the full counting statistics is manifestly
non-Poissonian - standard deviation becomes comparable to the average
number of particles - which remains true in other temperature regimes as well.

\section{Imaginary-time evolution of harmonic oscillator under sudden change
  of frequency and external force.}\label{sec:harmonic}

For high temperatures the modification $H_0\to
H_\lambda(r)$ amounts to a sudden change of the oscillator frequency 
$\omega_0 \to\omega_1 =\omega_0\sqrt{1 +2 s}$, where $s=\lambda \bar{n}
\ell_\varphi$ and $\omega_0 = 1/\bar{n}\ell_\varphi$.
We define dimensionless imaginary time $t_R=\bar{n} R$ and calculate 
the evolution of the ground state, $\Phi_\lambda(r;R)=\langle r|{e^{-t_R
    H_\lambda}}|0\rangle$ by adopting  the methods of  
Ref.~\cite{BazZeldovichPerelomov1} to the imaginary time evolution, with the
result
\begin{align}
\label{eq:evolution}
\Phi_\lambda(r;R) = \frac{1}{\sqrt{\pi} b} 
  e^{-\frac{ r^2}{2 }
  \left(\frac{1}{b^2}+\frac{1}{\omega_0}\frac{\dot{b} }{ b}\right)
  -\omega_0\int_0^{t_R} \frac{\dd \tau}{b^2(\tau)}} \, .
\end{align}
Here the scaling factor $b=b(t_R)$  is obtained from the solution of the
second order differential equation 
 \begin{align}
   \label{eq:bdiffeq}
-\ddot{b} +\omega_1^2 b =\frac{\omega_0^2}{ b^3}\, ,\qquad
b(0) =1,\qquad  \dot{b}(0)=0\, .
 \end{align}
Solving Eq.~(\ref{eq:bdiffeq}) we obtain
\begin{align}
  \label{eq:bres}
b(\tau) =\sqrt{1+\frac{\omega_1^2-\omega_0^2}{\omega_1^2} \sinh^2
  \omega_1\tau}\, ,
\end{align}
and 
\begin{align}
  \omega_0\int_0^{t_R} \frac{\dd
  \tau}{b^2(\tau)} = \frac{1}{2}
  \log\frac{1+\frac{\omega_0}{\omega_1} \tanh \omega_1 t_R}{1-\frac{\omega_0}{\omega_1} \tanh \omega_1 t_R}\, . 
\end{align}
Substituting these results into Eq.~(\ref{eq:evolution}) for the evolved
ground state  and calculating its
overlap with the initial state  yields  the generating function 
\begin{align}
  \label{eq:chiresfree}
\chi(\lambda,R) &= \frac{2 e^{-\omega_0\int_0^{t_R} \frac{\dd
  \tau}{b^2(\tau)}}}{b \,e^{-\omega_0 t_R}} \int_0^\infty\! r\dd r\,  e^{-\frac{ r^2}{2 }
  \left(\frac{1+b^2}{b^2}+\frac{1}{\omega_0}\frac{\dot{b} }{ b}\right)}
                  \nonumber\\
  &=\frac{2e^{\omega_0 t_R}e^{-\omega_0\int_0^{t_R} \frac{\dd
  \tau}{b^2(\tau)}}}{b+\frac{1}{b}+\frac{\dot{b}}{\omega_0}}\nonumber \\     
&=\frac{e^{\omega_0 t_R}
  \left(\cosh \omega_1 t_R -\frac{\omega_0}{\omega_1}\sinh\omega_1 t_R\right)}
  {1+\frac{\omega_1^2-\omega_0^2}{2\omega_0\omega_1} \sinh \omega_1t_R
    \left(\cosh\omega_1t_R+\frac{\omega_0}{\omega_1}\sinh\omega_1t_R\right)}\nonumber \\
                & = \frac{e^{R/\ell_\varphi}}{\cosh\sqrt{1+2s}\frac{ R}{\ell_\varphi}
                  + (1+s) \frac{\sinh\sqrt{1+2s}\frac{R}{\ell_\varphi}}{\sqrt{1+2s}}}
  \,.\end{align}
valid in the high-temperature regime for an interval of arbitrary length $R$.
The latter has to be compared to the microscopic length $\ell_\varphi$.
In the limit $R\ll \ell_\varphi$ we recover the
generating function 
\begin{align}
  \label{eq:chismallR}
\chi(\lambda,R) \simeq \frac{1}{1+\frac{\omega_1^2-\omega_0^2}{2\omega_0}t_R} = 
\frac{1}{1+\lambda\bar{n} R}\, ,
\end{align}
\ie $\chi(k/\bar{n}R,R) =1/(1+k)$ is  Laplace transform of the exponential
probability density, Eq. (10).
For long intervals, $R\gg \ell_\varphi$ we have 
\begin{align}
  \label{eq:chilargeR}
  \chi(\lambda,R) &\simeq
                &= \frac{4\sqrt{1+2s}}{\left(1+\sqrt{1+2s}\right)^2}
  e^{-\left(\sqrt{1+2s} -1\right)\frac{R}{\ell_\varphi}}\
\end{align}
leading to Eq.~(19). 


In the low temperature regime, $\ell_\varphi/\xi \gg 1$ one uses the expansion
(11) together with with the approximation 
\begin{align}
  \lambda r^2=\lambda(1+\delta r)^2 \simeq \lambda  +
  2\lambda  \delta r 
\end{align}
mapping the problem on a 2D harmonic oscillator under influence
of a time dependent force $f(\tau) = -2\lambda $ acting 
for $0<\tau<t_R$. Again, the
evolution of the ground state can be found adapting methods of
Ref.~\cite{BazZeldovichPerelomov1} to imaginary time and is given by
\begin{align}
  \label{eq:evoluforce}
  \Phi_\lambda(1+\delta r;R) = e^{-F(\delta r;t_R)-\lambda t_R} \Psi_0
  \left(\delta r -\eta \right)\, ,
\end{align}
where $\Psi_0(\delta r) =
(\ell_\varphi/2\pi^3\xi)^{1/4}e^{-(\ell_\varphi/\xi)\delta r^2}$ and 
$\eta = \eta(\tau)$ is the solution of the classical equation of motion 
\begin{align}
  \label{eq:classicaleta}
  \ddot{\eta} -\omega^2\eta = -\frac{f}{M} = \, 
\end{align}
with initial conditions $\eta(0) =\dot{\eta}(0) =0$ and 
\begin{align}
  F(\delta r; t_R)  &= M\dot{\eta} \left(\delta r-\eta\right) + 
  \int_0^{t_R}\! \dd \tau \left(\frac{M \dot{\eta}^2}{2} +\frac{M\omega^2\eta^2}{2}
  -f\eta\right) \nonumber\\
                  &=M\dot{\eta} \left(\delta r-\eta/2\right)-
  \frac{1}{2}\int_0^{t_R}\! \dd \tau \,f\eta\, ,  
\end{align}
where we have used equation of motion, Eq.~(\ref{eq:classicaleta}) to simplify
the integral.  Solving Eq.~(\ref{eq:classicaleta}) we get 
\begin{align}
  \eta(\tau) = \frac{f}{M\omega^2}\left(1-\cosh\omega \tau\right)\, .
\end{align}
Substituting it into Eq.~(\ref{eq:evoluforce})  and  truncating
at the second order in $\lambda$ in the exponent we get
\begin{align}
  \chi(\lambda,R) \simeq  \exp\left[-\lambda \bar{n} R+
  \frac{ C(R/\xi) \xi }{2 \ell_\varphi }(\lambda \bar{n} R)^2 
  \right] \,  ,
\end{align}
where the crossover function 
\begin{align}
  \label{eq:crossoverC}
  C(x) = \frac{1}{x^2}\left(x-\frac{1}{2}\left(1-e^{-2x}\right)\right)\, .
\end{align}
Performing
the inverse Laplace transform we get Eq.~(21) of the main text.


\section{Third and fourth moments.}

\begin{figure}[H]
\includegraphics[width=\columnwidth]{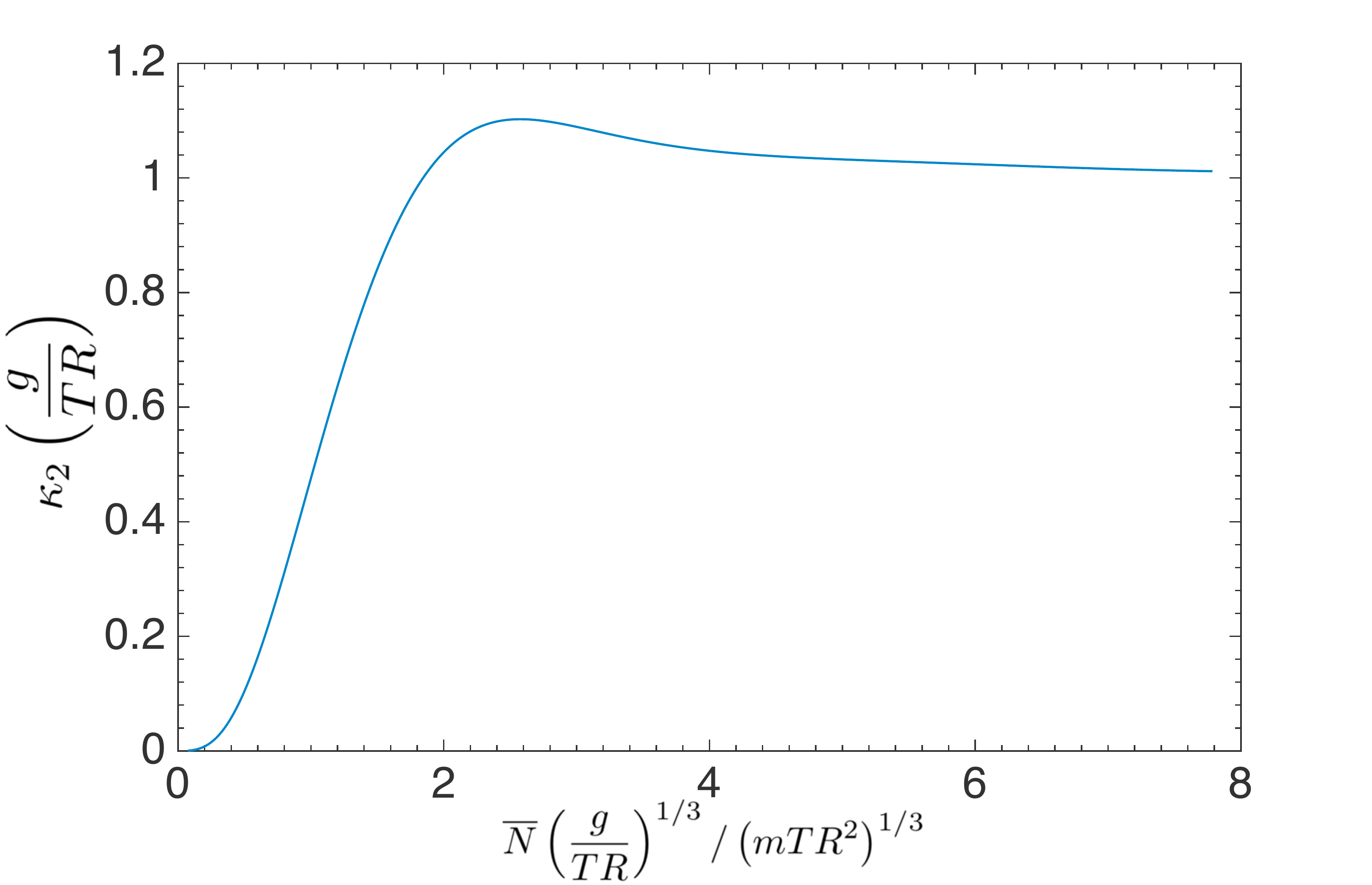}
\caption{The second cumulant, $\kappa_2=\overline{\delta N^2}$, of the large-R distribution as a function of $\overline{N}$ at fixed $R$ and $T$. Dependence on dimensionless parameters $TR/g$ and $mTR^2$ is shown for generality.
\label{fig:deltaN2vsN}}
\end{figure}
\begin{figure}[H]
\includegraphics[width=\columnwidth]{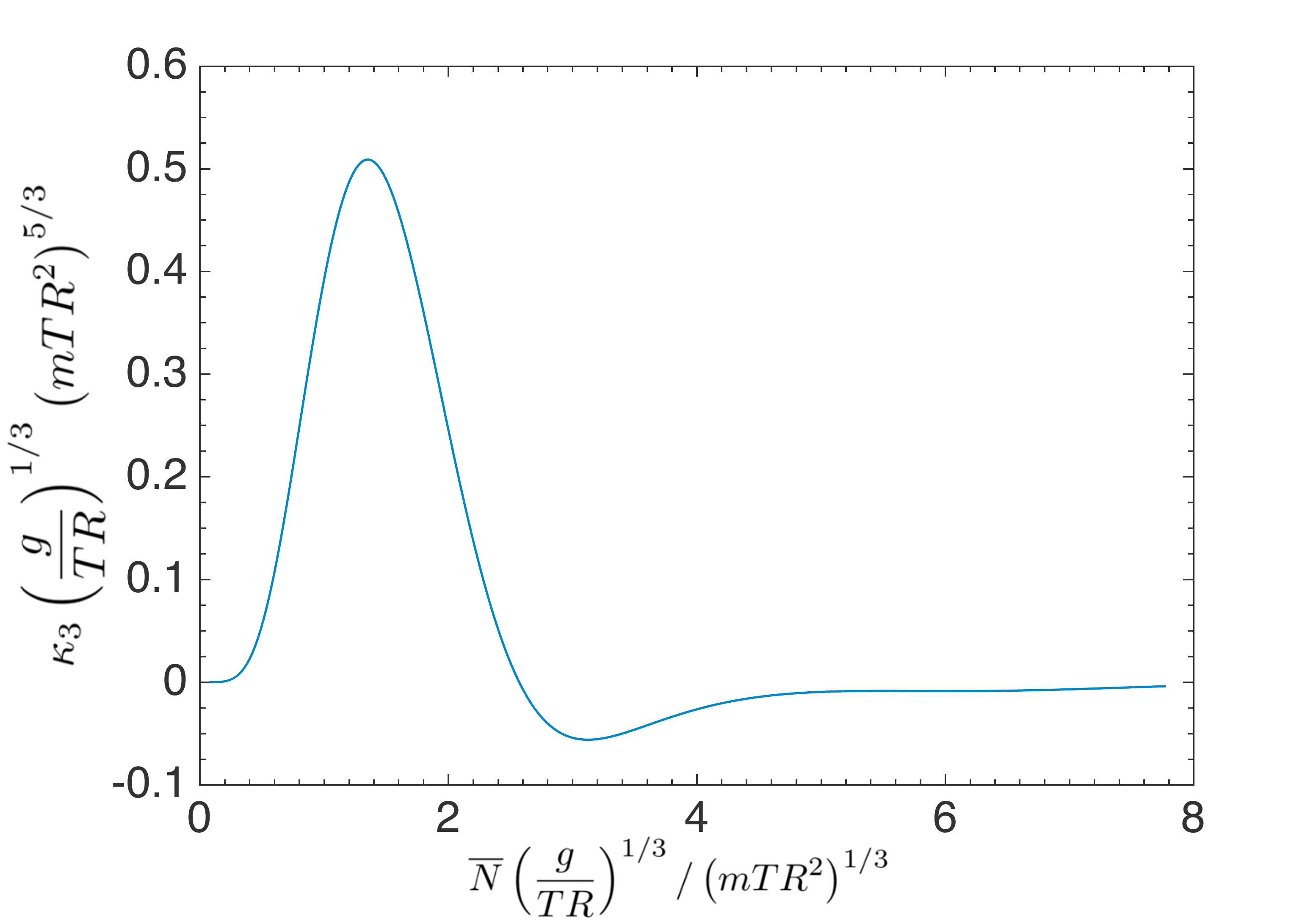}
\caption{The third cumulant, $\kappa_3=\overline{\delta N^3}$, of the large-R distribution as a function of $\overline{N}$ at fixed $R$ and $T$. Dependence on dimensionless parameters $TR/g$ and $mTR^2$ is shown for generality.
  \label{fig:deltaN3vsN}}

\end{figure}
\begin{figure}[H]
\includegraphics[width=\columnwidth]{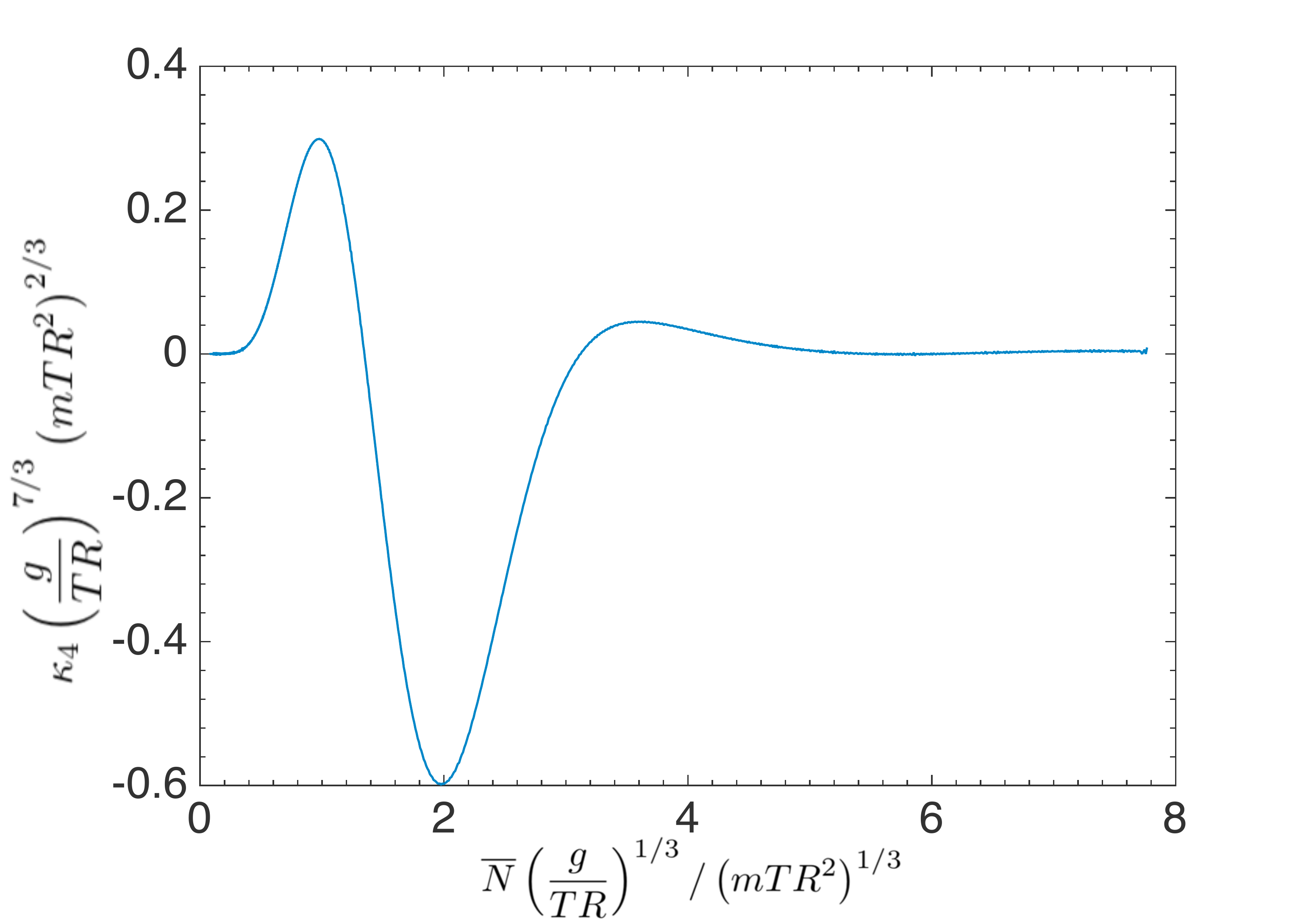}
\caption{The fourth cumulant, $\kappa_4$, of the large-R distribution as a function of $\overline{N}$ at fixed $R$ and $T$. Dependence on dimensionless parameters $TR/g$ and $mTR^2$ is shown for generality.
\label{fig:deltaN4vsN}}
\end{figure}
Similarly to expressing variance as a function of $T/c\bar{n}$, Eq.~(22),
second, third, and fourth cumulants can be plotted as functions of
$\overline{N}$ at fixed $R$ and $T$ to be  compared with the existing results obtained
using Yang-Yang thermodynamics in Ref.~\cite{ArmijoJacqmin1}. Defining
$\overline{\delta N^3}=\overline{\left(N-\overline{N}\right)^3}$ and
$\overline{\delta N^4}=\overline{\left(N-\overline{N}\right)^4}$ we present 
second, third and fourth cumulants, $\kappa_2=\overline{\delta N^2}$,
$\kappa_3=\overline{\delta N^3}$ and
$\kappa_4=\overline{\delta N^4}-3\left(\overline{\delta N^2}\right)^2$, as
functions of $\overline{N}$ in
Figs.~\ref{fig:deltaN2vsN},~\ref{fig:deltaN3vsN} and \ref{fig:deltaN4vsN}.
Notice that cumulants are expressed in terms of
universal functions which should be scaled appropriately depending on the
values of dimensionless parameters
$TR/g=\left(\bar{n}\xi\right)^2R/\ell_\varphi$ and
$mTR^2=\bar{n}R^2/\ell_\varphi$. Shapes shown on Figs.~\ref{fig:deltaN2vsN}
and \ref{fig:deltaN3vsN} are in full agreement  to  Fig. 2(b) of \cite{ArmijoJacqmin1}.

\end{document}